\documentclass[]{raa}            % referee version: for submission
\usepackage{graphicx,times}
\usepackage{natbib}
\usepackage{amssymb}
\usepackage{ulem}
\usepackage{color}

\providecommand{\bm}[1]{\mbox{\boldmath$#1$\unboldmath}}

\begin{document}

   \title{Discovery of a deep, low mass ratio overcontact binary GSC 03517-00663 }

 \volnopage{ {\bf 2014} Vol.\ {\bf X} No. {\bf XX}, 000--000}
   \setcounter{page}{1}

   \author{ Guo, D.-F.
   \inst{1,2}
      \and Li, K.
      \inst{1,2,3}
      \and Hu, S.-M.
      \inst{1,2}
      \and Jiang, Y.-G.
      \inst{1,2}
      \and Gao, D.-Y.
      \inst{1,2}
      \and Chen, X.
      \inst{1,2}
   }

   \institute{Institute of Space Sciences, School of Space Science and Physics, Shandong University, Weihai, 264209, China; {\it e-mail: kaili@sdu.edu.cn, likai@ynao.ac.cn (Li, K.); husm@sdu.edu.cn (Hu, S.-M.)} \\
        \and
        Shandong Provincial Key Laboratory of Optical Astronomy and Solar-Terrestrial Environment, Weihai, 264209, China
        \and
             Key Laboratory for the Structure and Evolution of Celestial
Objects, Chinese Academy of Sciences
% \email{likai@ynao.ac.cn}
\vs \no
%  {\small Received [year] [month] [day]; accepted [year] [month] [day] }
}

\abstract{When observing the blazars, we identified a new eclipsing binary GSC 03517-00663. The light curves of GSC 03517-00663 are typical EW-type light curves. Based on the observation using the 1m telescope at Weihai Observatory of Shandong University, complete $VRI$ light curves were determined. Then, we analyzed the multiple light curves using the W-D program. It is found that GSC 03517-00663 has a mass ratio of $q=0.164$ and a contact degree of $f=69.2\%$. GSC 03517-00663 is a deep, low mass ratio overcontact binary. The light curves of GSC 03517-00663 show strong O'Connell effect, it was explained by employing a dark spot on the secondary component.
\keywords{stars: binaries: close --- stars: binaries: eclipsing ---
stars: individual(GSC 03517-00663)} }

   \authorrunning{Guo et al. }    %author_head in even pages
   \titlerunning{ Discovery of a deep, low mass ratio overcontact binary GSC 03517-00663 }  % title_head in odd pages
   \maketitle
\section{Introduction}           %% first-level sections will be auto-capitalized
\label{sect:intro}
W UMa type stars are usually contact binaries, both component stars are in contact with each other and sharing a common convective envelope. This type of binaries are usually composed of two cool, main-sequence stars with spectral types of F to K and they normally show typical EW-type light curves, where light variation is continuous and has a very small difference between the depths of the two minima. The nearly equal depths of the two
minima reveal that the effective temperatures of both components are very similar despite different component masses (\citealt{qia14}).

Deep, low mass ratio overcontact binary is a system which has a mass ratio of $q < 0.25$ and the overcontact degree of $f > 50\%$. This kind of binaries is at the late evolutionary stage. The final stages of the evolution of them are still not well understood. There are several clues that they can be the progenitors of rapidly-rotating single giants (Blue Stragglers and FK Com type stars) by
coalescence of two components through continuing angular momentum loss (\citealt{egg01}). More observation and investigation of this type of binaries are needed.

In this paper, we presented $VRI$ light curves of a W UMa type binary GSC 03517-00663 ($\alpha_{2000}=17^{h}28^{m}55.14^{s}$, $\delta_{2000}=+50^{\circ}16^{\prime}17.5^{\prime\prime}$). GSC 03517-00663 is a new binary discovered  by us when we observed blazars. This paper follows the following structure. CCD observations, discovery and orbital period determination of GSC 03517-00663 are shown in Section 2. In Section 3, light curves investigation is presented. The conclusions are discussed in Section 4.

\section{Observation, discovery and period determination}
\label{sect:obser}
While analyzing the observational data directed toward the study of blazars (OT546), we found that one of the field stars varies more than 0.4 magnitudes ($V$ band) during one night. Neither the GCVS nor the NSV catalogues contain this star, so we conclude that it is a newly identified variable star. According to the GSC 1.2, this star is named GSC 03517-00663.

CCD photometric observations of GSC 03517-00663 were carried out in May and June, 2009 and October, 2012, using a PIXIS 2048B CCD camera attached to the 1.0 m Cassegrain telescope (\citealt{hu14}) at Weihai Observatory of Shandong University. The PIXIS camera has $2048\times2048$ square pixels (13.5$\times$13.5$\mu$m pixel$^{-1}$), providing an effective field of view about 11.8$'$ $\times$ 11.8$'$. The standard Johnson and Cousins filters ($V$, $R$, and $I$) were used during our observations. The typical integration times for each image were 200s, 150s and 120s in $V$, $R$ and $I$ bands, respectively. The reductions of observations were conducted using the APPHOT packages in IRAF\nolinebreak\footnotemark[1] procedures.\footnotetext[1]{IRAF is distributed by the National Optical Astronomy Observatories, which is operated by the Association of Universities for Research in Astronomy Inc., under contract to the National Science Foundation.} All data were processed by bias and flat-field correction. One of the CCD images is shown in Figure 1, where ``V'' refers to the variable star (i.e., GSC 03517-00663 ), ``C'' to the comparison star, and ``CH'' to the check star. Standard stars  B  ( $\alpha_{2000.0}=17^h28^m24^s.6$, $\delta_{2000.0}=50^{\circ} 14^{\prime}35$\arcsec$ .6$ )  and  H ($\alpha_{2000.0}=17^h28^m14^s.3$, $\delta_{2000.0}=50^{\circ} 12^{\prime}40$\arcsec$ .2$ ) taken from \cite{fiorucci06} were used as the comparison star and check star, respectively.

\begin{figure*}
\begin{center}
\includegraphics[angle=0,scale=0.6]{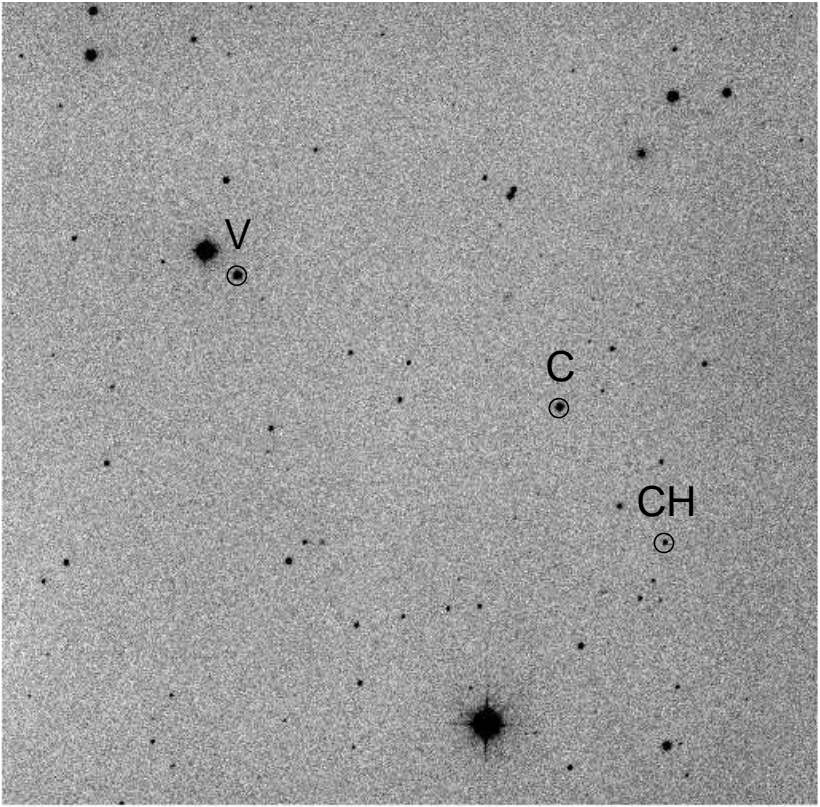}
\caption{CCD image in the field of view around GSC 03517-00663. "V" refers to the variable star (i.e., GSC 03517-00663 ), "C" to the comparison star, and "CH" to the check star. }
\end{center}
\end{figure*}

Jurkevich method (\citealt{jur71}) was applied to all the $V$ band data for periodicity analysis. Jurkevich method is based on the expected mean square
deviation and the unequally spaced observations, so it is less
inclined to generate a spurious periodicity comparing with a Fourier analysis.
It involves testing a series of trial periods and the data are
folded according to the trial periods. According to their phases around each trial period, all data are divided
into $m$ groups. The variance $V_{i}^{2}$
for each group and the sum of each group variance $V_{m}^{2}$  are
computed. If a trial period equals to the real one, $V_{m}^{2}$
would reach its minimum. The results derived by the Jurkevich method using  m=50 are shown in Figure 2. The minimum value indicates the period of 0.295025 days.

\begin{figure*}
\begin{center}
\includegraphics{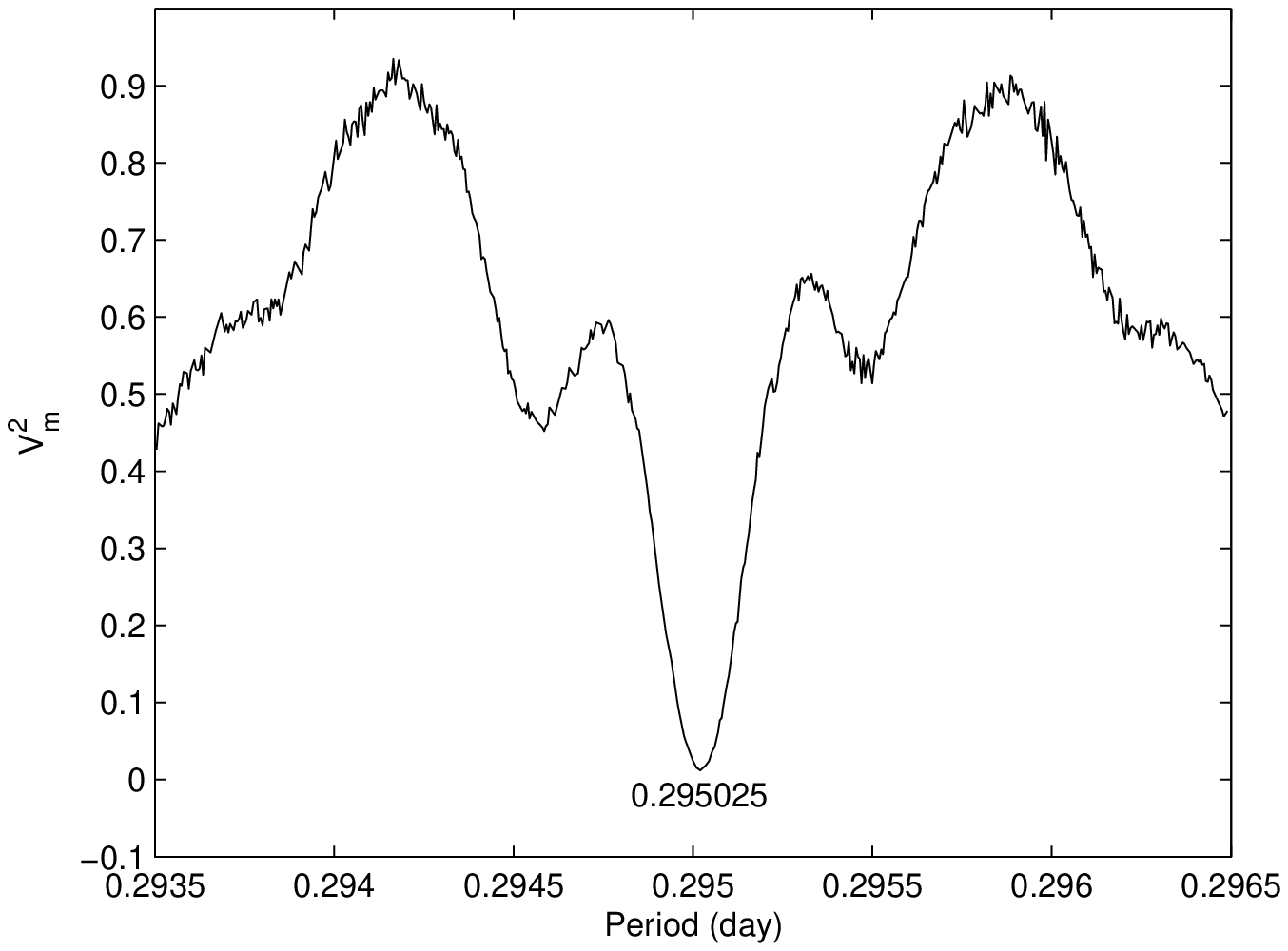}

\caption{Relationship between the trial period and $V_{m}^{2}$ using all the $V$ band data.}
\end{center}
\end{figure*}

The observed $VRI$ band light curve folded according the period of 0.295025 days is shown in Figure 3. From this figure, it is seen that the data
observed in 2009 merged smoothly and the light variation is of W UMa type eclipsing binary. Ten times of light minimum were determined and are listed in Table 1. Using the following ephemeris
\begin{eqnarray}
Min.I = HJD2454974.2478 + 0.295025E,
\end{eqnarray}
the $O-C$ values are calculated and are listed in Table 1. Then, through a least squares solution, a new linear ephemeris is determined from all these data,
\begin{eqnarray}
Min.I = HJD2454974.24763\pm0.00045 + 0.29502416\pm0.00000034E.
\end{eqnarray}
\begin{table}
\begin{center}
\caption{New determined times of light minimum for GSC 03517-00663}
\begin{tabular}{lcccr}
\hline
  JD(Hel.) &     Errors &       Min. & E & O-C \\\hline
2454974.2478 & $\pm$0.0003 & p & 0      & 0.0000  \\
2454975.2797 & $\pm$0.0004 & s & 3.5    & -0.0007 \\
2454977.1995 & $\pm$0.0017 & p & 10     & 0.0015  \\
2454978.2289 & $\pm$0.0004 & s & 13.5   & -0.0017 \\
2454979.2646 & $\pm$0.0003 & p & 17     & 0.0014  \\
2454982.2136 & $\pm$0.0003 & p & 27     & 0.0001  \\
2455003.1605 & $\pm$0.0003 & p & 98     & 0.0003  \\
2455005.0752 & $\pm$0.0005 & s & 104.5  & -0.0027 \\
2455005.2255 & $\pm$0.0006 & p & 105    & 0.0001  \\
2456207.0061 & $\pm$0.0002 & s & 4178.5 & -0.0037 \\

\hline

\end{tabular}
\end{center}
\end{table}

\begin{figure*}
\begin{center}
\includegraphics{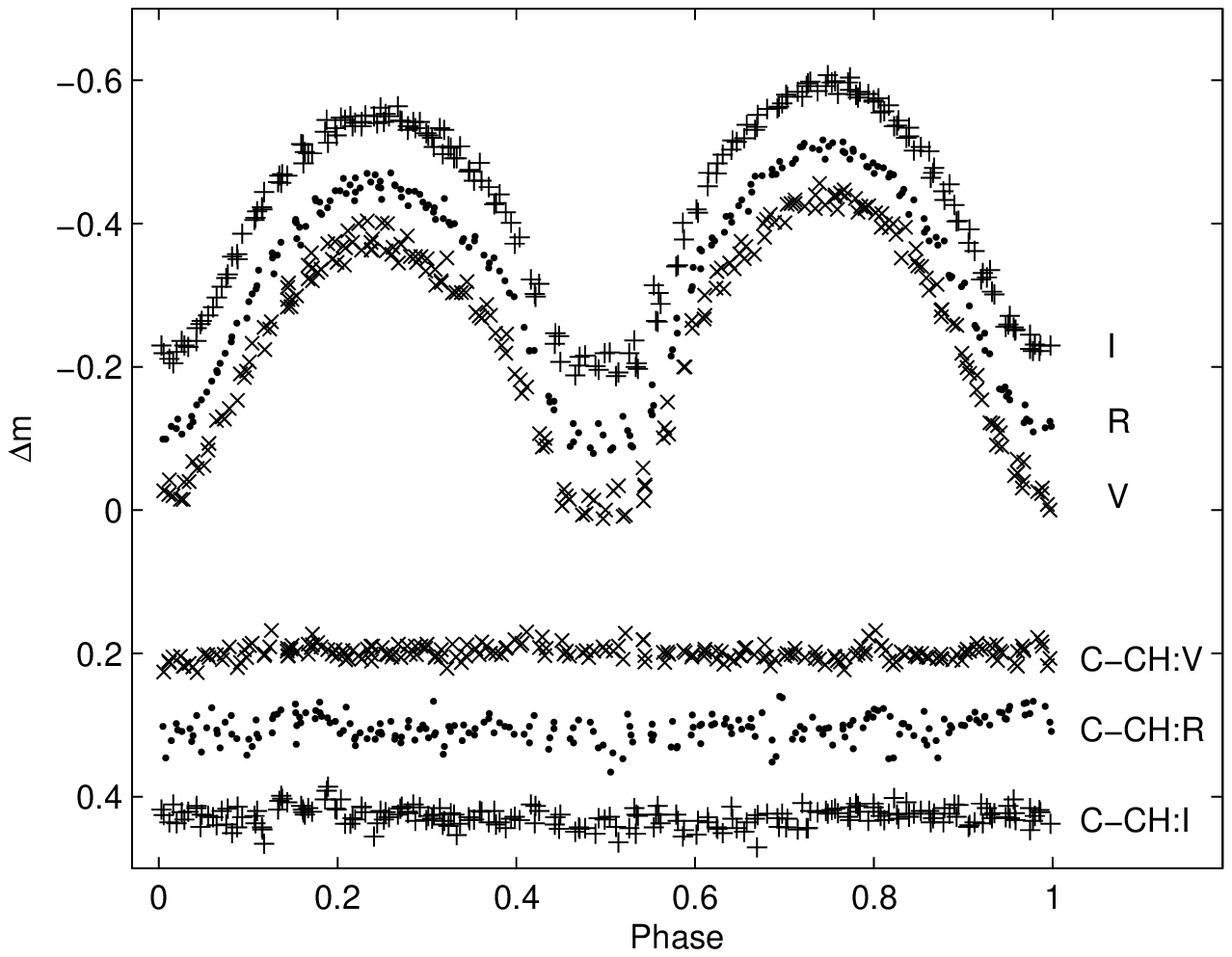}
\caption{$VRI$ light curves of GSC 03517-00663 observed in 2009. The phases were calculated using
equation (1). Different symbols represent
different bands.}
\end{center}
\end{figure*}

\section{Light curve analysis}
\label{sect:Light}
Using the fourth version of the W-D program (\citealt{wil71,wil90,wil94,wil03}), we analyzed the $V$, $R$ and $I$ light curves of GSC 03517-00663. According to NOMAD (The Naval Observatory Merged Astrometric Dataset, \citealt{zac04}), the color index of GSC 03517-00663 can be derived to be $B-V=0.62$, which is corresponding to the spectral type of G2 due to \cite{cox00}. The gravity darkening coefficients of the two components were taken to be $g_1=g_2=0.32$ for convective atmosphere from \cite{luc67}. The bolometric albedo coefficients of the two components were fixed at $A_1=A_2=0.5$ for convective atmospheres following \cite{ruc69}. The bolometric and bandpass limb-darkening coefficients of the two components were taken from \cite{van93}.

Starting with the solutions by mode 2, we found that the solutions are usually converged when both components fill their Roche lobes. So, the final iterations were made in mode 3, which corresponds to contact configuration. The quantities varied in the solutions were the mass ratio $q$, the effective temperature of the secondary component $T_2$, the monochromatic luminosity of primary component in $V$, $R$ and $I$ bands $L_1$, the orbital inclination $i$ and the dimensionless potential of the primary component $\Omega_1$ ($\Omega_1=\Omega_2$). As an obvious O'Connell effect can be seen in the light curves of GSC 03517-00663, the solutions were calculated on two cases: with or without spot. In case one, a solution with one dark spot on the secondary component leads a good fit to the light curves.
Since GSC 03517-00663 is a newly discovered binary, no mass ratio has been obtained. A $q$-search method was used to determine the mass ratio. Solutions were carried out for a series of values of the mass ratio (from 0.123 to 1.0). The relation between the resulting sum $\Sigma$ of weighted square deviations and $q$ is plotted in Figure 4. The minimum value was obtained at $q=0.16$. Then, we chose $q=0.16$ as a initial value and made it an adjusted parameter. When the solution converged, the result was determined. The solutions based on the two cases are listed in Table 2. The residual of the solution with spot is much smaller than that without spot. Therefore, we adopted case one as the final solution. The comparison between observed and the theoretical light curves is shown in Figure 5. Figure 6 shows the configuration of this system at phase 0.25.

\begin{table}
\begin{center}
\caption{ Photometric solutions for GSC 03517-00663}
\begin{tabular}{lclcl}
\hline
Parameters & Without Spot & Errors & With Spot & Errors  \\

\hline
    $ g_1=g_2$ &  0.32 & Assumed& 0.32 &Assumed \\
     $A_1=A_2$ &  0.5 & Assumed&  0.5 & Assumed\\
     $x_{1bol},$ $x_{2bol}$& 0.648, 0.647& Assumed& 0.648, 0.647& Assumed\\

     $y_{1bol},$ $y_{2bol}$& 0.207, 0.221& Assumed& 0.207, 0.221& Assumed\\

     $x_{1V},$ $x_{2V}$& 0.762, 0.745& Assumed& 0.762, 0.745& Assumed\\

     $y_{1V},$ $y_{2V}$& 0.232, 0.256& Assumed& 0.232, 0.256& Assumed\\

     $x_{1R},$ $x_{2R}$& 0.670, 0.653& Assumed& 0.670, 0.653&Assumed\\

     $y_{1R},$ $y_{2R}$& 0.250, 0.267& Assumed& 0.250, 0.267& Assumed\\

     $x_{1I},$ $x_{2I}$& 0.576, 0.560& Assumed& 0.576, 0.5604&Assumed\\

     $y_{1I},$ $y_{2I}$& 0.244, 0.256& Assumed&0.244, 0.256& Assumed\\

    $ T_1(K) $&  5800&     Assumed  &  5800&     Assumed   \\

  $q(M_2/M_1) $& 0.162  & $ \pm0.003$  & 0.164 &$ \pm0.002$   \\

  $ \Omega_{in} $ & 2.1344 & Assumed &2.1416 &Assumed \\

  $ \Omega_{out} $ & 2.0303& Assumed &2.0359 &Assumed \\

        $T_2(K) $&       6024 & $ \pm29$&       6075 & $ \pm18$\\

         $i$ &     77.125& $ \pm0.733$&     77.446 & $ \pm0.484$\\

$L_{1V}/L_V$&      0.8013 & $ \pm0.0008$& 0.7878 & $ \pm0.0005$\\
$L_{1R}/L_R$&      0.8068 & $ \pm0.0007$& 0.7947 & $ \pm0.0004$\\
$L_{1I}/L_I$&      0.8107 & $ \pm0.0006$& 0.7997 & $ \pm0.0004$\\

  $\Omega_1$=$\Omega_2$ & 2.0803 & $ \pm0.0099$& 2.0685 & $ \pm0.0070$\\

  $r_1(pole)$ &   0.5164 &$ \pm 0.0027$&   0.5201 &$ \pm 0.0019$\\

  $r_1(side)$ &   0.5713 & $ \pm0.0042$&   0.5774 & $ \pm0.0031$\\

  $r_1(back)$ &     0.5968 & $ \pm0.0053$& 0.6047 & $ \pm0.0039$\\

  $r_2(pole)$ &     0.2351 & $ \pm0.0065$& 0.2422 & $ \pm0.0044$\\

  $r_2(side)$ &   0.2468 & $ \pm0.0080$&   0.2554 & $ \pm0.0056$\\

  $r_2(back)$ &   0.2993 & $ \pm0.0215$&   0.3203 & $ \pm0.0186$\\
  $f$&52.0\% & $\pm9.5\%$& 69.2\%&$\pm6.7\%$\\
  $\theta(radian)$ &-  &- &  1.406    &$ \pm0.290$\\
  $\phi(radian)$  &-  &- & 1.593     &$ \pm0.071$ \\
  $r(radian)$ &-  &- &  0.484    &$ \pm0.046$ \\
  $T_f(T_d/T_0)$ &-  &- &  0.647    &$ \pm0.086$ \\
  $\Sigma W(O-C)^2$ &   0.0144 &  &   0.0055 &  \\
\hline
\end{tabular}
\end{center}
\end{table}

\begin{figure}
\begin{center}
\includegraphics[angle=0,scale=1.0]{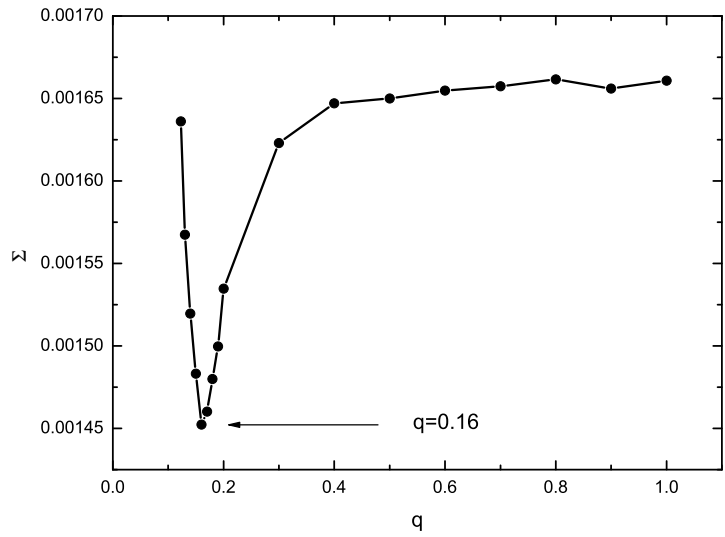}
\caption{Relation between $\sum$ and q for GSC 03517-00663.}
\end{center}
\end{figure}
\begin{figure}
\begin{center}
\includegraphics[angle=0,scale=1.0]{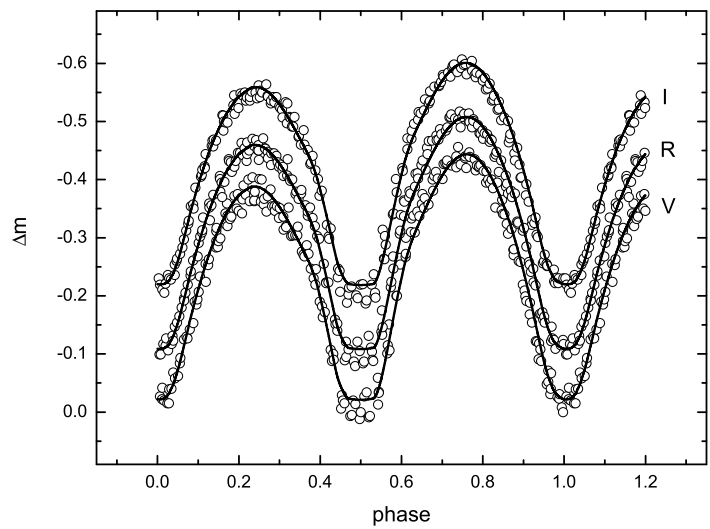}
\caption{Observed (open circles) and theoretical (solid lines) light curves of GSC 03517-00663.}
\end{center}
\end{figure}
\begin{figure}
\begin{center}
\includegraphics[angle=0,scale=1.0]{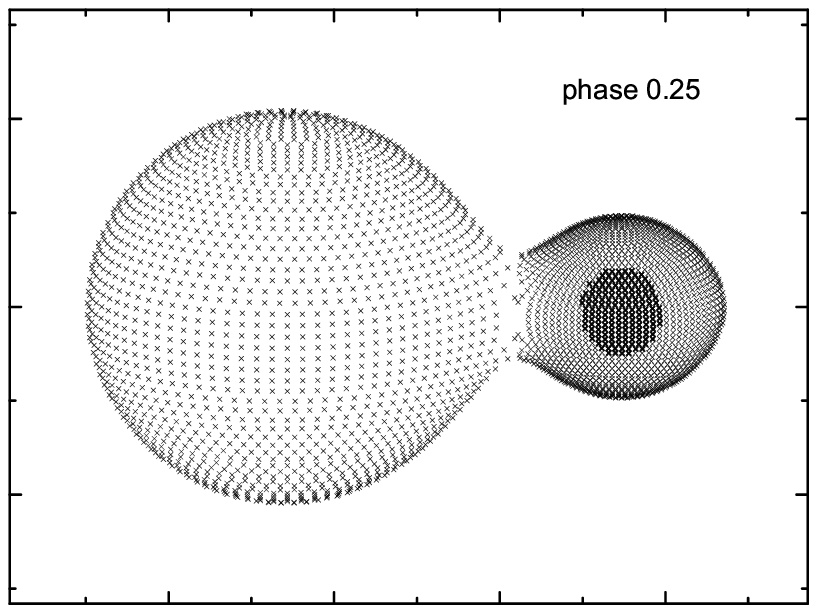}
\caption{Configuration of the low mass ratio, high fill-out overcontact binary GSC 03517-00663 at phase 0.25.}
\end{center}
\end{figure}

\section{Results and Discussions }
In this paper, we presented a newly discovered W UMa binary GSC 03517-00663. Using the Jurkevich method, the orbital period of GSC 03517-00663 was determined to be $P=0.295025$ days. Based on the $V$, $R$ and $I$ light curves, photometric solutions for the newly discovered eclipsing binary GSC 03517-00663 have been derived. We found that the mass ratio of GSC 03517-00663 is $q=0.164$ and that the contact degree, defined by $f=(\Omega_{in}-\Omega)/(\Omega_{in}-\Omega_{out})$, is $f = 69.2\%$.
As V857 Her, QX And, EM Pis and XY Leo (\citealt{qia05,qia07,qia08,qia11}), GSC 03517-00663 is a deep, low mass ratio overcontact binary.

The spectral type of GSC 03517-00663 is G2, it is a solar like binary system. The deep convective envelope along with fast rotation can produce strong magnetic activity. Therefore, the disagreement between the two maxima of the light curves was explained by the presence of a dark spot on the common convective envelope of the secondary component.

GSC 03517-00663 is a solar like deep, low mass ratio overcontact binary. It may be the progenitor of Blue Straggler/FK Com-type stars. Future observations are needed to determine the evolution of the binary and to analyze the orbital period variation.

\normalem
\begin{acknowledgements}
This work is partly supported by the National Natural Science Foundation of China (Nos.
11203016, 11333002, 10778619, 10778701, U1431105), and by the
Natural Science Foundation of Shandong Province (No. ZR2012AQ008) and by the Open Research Program of Key Laboratory for the Structure and Evolution of Celestial Objects (No. OP201303). Thanks the anonymous referee very much for her/his very positive comments and helpful suggestions.

\end{acknowledgements}

\label{lastpage}


\begin{thebibliography}{99}
\bibitem[Cox (2000)]{cox00} Cox A. N., 2000, Allen's astrophysical quantities, 4th ed. Publisher: New York: AIP Press; Springer
\bibitem[Eggleton \& Kiseleva-Eggleton (2001)]{egg01} Eggleton, P., \& Kiseleva-Eggleton, L., 2001, ApJ, 562, 101
\bibitem[Fiorucci \& Tosti (1996)]{fiorucci06} Fiorucci, M., \& Tosti, G 1996, A\&AS, 117,457
\bibitem[Hu et al. (2014)]{hu14} Hu, S. M., Han, S. H., Guo, D. F., Du, J. J., 2014, RAA, 14, 719
\bibitem[Jurkevich (1971)]{jur71} Jurkevich, I., 1971, Ap\&SS,  13, 154
\bibitem[Lucy (1967)]{luc67} Lucy, L. B., 1967, Z. Astrophys., 65, 89
\bibitem[Qian et al. (2008)]{qia08} Qian, S.-B., He, J.-J., Soonthornthum, B., Liu, L., Zhu, L.-Y., Li, L.-J., Liao, W. P., Dai, Z.-B., 2008, AJ, 136, 1940
\bibitem[Qian et al. (2011)]{qia11} Qian, S.-B., Liu, L., Zhu, L.-Y., He, J.-J., Yang, Y.-G., Bernasconi, L., 2011, AJ, 141, 151
\bibitem[Qian et al. (2007)]{qia07} Qian, S.-B., Liu, L., Soonthornthum, B., Zhu, L.-Y., He, J.-J., 2007, AJ, 134, 1475
\bibitem[Qian et al. (2014)]{qia14} Qian, S.-B., Wang, J.-J., Zhu, L.-Y., Snoonthornthum, B., Wang, L.-Z., Zhao, E. G., Zhou, X., Liao, W.-P., Liu, N.-P., 2014, ApJS, 212, 4
\bibitem[Qian et al. (2005)]{qia05} Qian, S.-B., Zhu, L.-Y., Soonthornthum, B., Yuan, J.-Z., Yang, Y.-G., He, J.-J., 2005, AJ, 130, 1206
\bibitem[Ruci\'{n}ski (1969)]{ruc69} Ruci\'{n}ski, S. M., 1969, Acta Astronomica, 19, 245
\bibitem[van Hamme (1993)]{van93} van Hamme, W., 1993, AJ, 106, 2096
\bibitem[Wilson (1990)]{wil90} Wilson, R. E., 1990, ApJ, 356, 613
\bibitem[Wilson (1994)]{wil94} Wilson, R. E., 1994, PASP, 106, 921
\bibitem[Wilson \& Devinney (1971)]{wil71} Wilson, R. E., \& Devinney, E. J., 1971, ApJ, 166, 605
\bibitem[Wilson \& Van Hamme (2003)]{wil03} Wilson, R. E., \& Van Hamme, W., 2003, Computing Binary Stars Observables, 4th edn of the W-D program, available at ftp.astro.ufl.edu/pub/wilson/lcdc2003
\bibitem[Zacharias et al. (2004)]{zac04} Zacharias, N., Monet, D. G., Levine, S. E., Urban, S. E., Gaume, R., Wycoff, G. L. 2004, AAS, 205, 4815



\end{thebibliography}
\end{document}